\documentclass[useAMS,usenatbib]{mnras}

\usepackage{newtxtext,newtxmath}
\usepackage[T1]{fontenc}
\usepackage{ae,aecompl}

\usepackage{graphicx}	
\usepackage{amsmath}	
\usepackage{amssymb}	
\usepackage{bm}
\usepackage{multirow}
\usepackage[dvipsnames,table]{xcolor} 
\usepackage{tikz}
\usetikzlibrary{positioning}

\newcommand{\us}[1]{{\mathrm{~#1}}} 
\renewcommand\u[1]{{\mathrm{#1}}} 
\DeclareMathOperator{\E}{\mathbb{E}} 
\newcommand{\elbo}{\operatorname{ELBO}}

\title[Differentiable Strong Lensing]{Differentiable Strong Lensing: Uniting Gravity and Neural Nets through Differentiable Probabilistic Programming}

\author[M. Chianese et al.]{
Marco Chianese,$^{1}$\thanks{m.chianese@uva.nl}
Adam Coogan,$^{1}$\thanks{a.m.coogan@uva.nl}
Paul Hofma$^{1}$\thanks{paul.hofma@student.uva.nl}
Sydney Otten$^{1,2}$\thanks{s.m.m.otten@uva.nl}
\newauthor
and Christoph Weniger$^{1}$\thanks{c.weniger@uva.nl}
\\
$^{1}$Gravitation Astroparticle Physics Amsterdam (GRAPPA),
Institute for Theoretical Physics Amsterdam
and Delta Institute for\\
~~Theoretical Physics,
University of Amsterdam, Science Park 904, 1098 XH Amsterdam, The Netherlands\\
$^{2}$Institute for Mathematics, Astrophysics and Particle Physics (IMAPP),
Radboud University, Heyendaalseweg 135,\\
~~6525 AJ Nijmegen, The Netherlands
}

\date{Accepted XXX. Received YYY; in original form ZZZ}

\pubyear{2019}

\begin{document}
\label{firstpage}
\pagerange{\pageref{firstpage}--\pageref{lastpage}}
\maketitle

\begin{abstract}
Since upcoming telescopes will observe thousands of strong lensing systems, creating fully-automated analysis pipelines for these images becomes increasingly important. In this work, we make a step towards that direction by developing the first end-to-end differentiable strong lensing pipeline. Our approach leverages and combines three important computer science developments: (a) convolutional neural networks, (b) efficient gradient-based sampling techniques, and (c) deep probabilistic programming languages.  The latter automatize parameter inference and enable the combination of generative deep neural networks and physics components in a single model. In the current work, we demonstrate that it is possible to combine a convolutional neural network trained on galaxy images as a source model with a fully-differentiable and exact implementation of gravitational lensing physics in a single probabilistic model. This does away with hyperparameter tuning for the source model, enables the simultaneous optimization of nearly one hundred source and lens parameters with gradient-based methods, and allows the use of efficient gradient-based posterior sampling techniques. These features make this automated inference pipeline potentially suitable for processing a large amount of data. By analyzing mock lensing systems with different signal-to-noise ratios, we show that lensing parameters are reconstructed with percent-level accuracy. More generally, we consider this work as one of the first steps in establishing \emph{differentiable probabilistic programming} techniques in the particle astrophysics community, which have the potential to significantly accelerate and improve many complex data analysis tasks.
\end{abstract}

\begin{keywords}
gravitational lensing: strong -- galaxies: structure -- dark matter.
\end{keywords}



\section{Introduction} 

Strong lensing is a gravitational effect through which an astrophysical light source is observed in distorted, multiple images in the sky due to the deflection of its light by matter distributed along the line of sight~\citep{Treu:2010uj}. It has become one of the main ways to probe the small-scale structure of dark matter halos, since subhalos with mass below $\sim 10^8~M_\odot$ do not host stars and are thus invisible~\citep{Fitts:2016usl,Read:2017lvq}. The detection (or non-detection) of these subhalos is a critical tool for discriminating among different paradigms of dark matter (DM)~\citep{Bertone:2004pz, Bertone:2018xtm}. In the standard $\Lambda \mathrm{CDM}$ cosmological model, the large-scale structures of the Universe form through the collapse of primordial density fluctuations. The matter content of the Universe is dominated by non-relativistic and almost collisionless substance dubbed cold dark matter (CDM). In this scenario, an abundance of small DM substructures is formed, as confirmed by ab-initio $N$-body cosmological simulations~\citep{Kuhlen:2012ft}. On the other hand, alternative well-motivated particle DM scenarios such as warm dark matter (WDM)~\citep{Bode:2000gq,Lovell:2013ola} and self-interacting DM~\citep{Vogelsberger:2018bok,Kahlhoefer:2019oyt} predict a lower abundance of low-mass DM substructures. Recent analyses of strong gravitational lensing of extended sources~\citep{2010MNRAS.408.1969V,Vegetti:2012mc,Hezaveh:2016ltk,Vegetti:2018dly,Ritondale:2018cvp} and quasars~\citep{2012MNRAS.419..936F,Gilman:2019vca,Gilman:2019nap} have already demonstrated sensitivity to DM haloes with masses larger than $10^8~M_\odot$. In the near future, new observatories like DES~\citep{Abbott:2016ktf}, LSST~\citep{2009arXiv0912.0201L,Drlica-Wagner:2019xan,2019arXiv190205141V}, Euclid~\citep{2010arXiv1001.0061R}, and next-generation observatories like ELT~\citep{Simon:2019kmm} will observe thousands of strong lensing systems with very high precision, pushing the sensitivity of lensing probes of DM substructures to even lower masses. Moreover, these observations will be dominated by lenses at high redshift, increasing the likelihood of detecting small DM haloes along the line-of-sight~\citep{Despali:2017ksx}.

In analyzing a galaxy-galaxy strong lens, the observed lensed image is reconstructed by simultaneously modeling the surface brightness of the source and the matter distribution of the lens galaxy. This requires parametrizing both of these components. While $N$-body simulations show that the density profiles of galactic DM halos are well-described by various analytic profiles~\citep{Navarro:1995iw}, the distribution of the source's light is more complicated. The Sersic brightness profile~\citep{Sersic} is a common choice (see e.g. \cite{10.1111/j.1365-2966.2010.18074.x}), but is inadequate for high-resolution observations and modeling high-redshift source galaxies, which generally have more complex morphologies than low-redshift ones. Another class of methods computes the source brightness profile on a grid by linearly inverting the observed lensed image given a fixed lens model~\citep{Warren:2003na,Suyu:2006fd}. This requires using specific prior with a particular form to regularize the source, depending on two-point quantities calculated between pairs of pixels as well as hyperparameters. These methods can be cast in a fully Bayesian framework and performed on an adaptive grid in the source plane~\citep{Koopmans:2005nr,Vegetti:2008eg}. The public code PyAutoLens implements this analysis strategy~\citep{Nightingale:2017cdh,pyautolens}. Extensions of these methods include grid-free approaches using radial basis functions centered on image pixels ray-traced back to the source plane~\citep{Merten:2014eia} and methods decomposing the source as a sum of shapelets~\citep{Birrer:2015rpa}. These methods are available in the SaWLens2~\citep{sawlens2} and Lenstronomy~\citep{Birrer:2018xgm,lenstronomy} software packages. Depending on the pipeline's fitting scheme, the choice of priors for the lensing system's parameters may be restricted, making it challenging to perform sensitivity analysis. These analysis pipelines also generally require dedicated, time-consuming hyperparameter optimization efforts for fitting each strong lensing system. On the other hand, fully-automated lensing pipelines which do not require these interventions will become increasingly useful for analyzing large strong lensing datasets, and for subsequently characterizing the underlying dark matter subhalo mass function. To this aim, it is of paramount importance to develop automated lens modeling techniques. In this paper, we demonstrate that this is possible through the use of new computational tools from the field of deep learning such as \emph{automatic differentiation} (AD). This enables the use of powerful gradient descent based methods which are crucial to speed up and automatize the fitting procedure of high-dimensional parameter space. In particular, this paper focuses on how deep generative models can be combined with known physics using AD to create a new pipeline for analyzing images of galaxy-galaxy strong lensing systems.

Deep learning has advanced dramatically over the past decade, with accurate image classifiers~\citep{AlexNet} and a host of generative methods such as variational autoencoders~\citep{2013arXiv1312.6114K,2014arXiv1401.4082J}, generative adversarial networks~\citep{GANOriginal,BigGAN,StyleGAN} and flow-based models~\citep{Glow} capable of producing novel, realistic images counting among its successes. These methods have also been applied to the physical sciences~\citep{Carleo:2019ptp}, with topics including deblending galaxy images~\citep{Reiman_2019}, generating weak gravitational lens convergence maps~\citep{CosmoGAN}, and classifying LHC jet events~\citep{Larkoski:2017jix}. However, there has been considerably less scientific investigation into leveraging automatic differentiation~\citep{DBLP:journals/corr/BaydinPR15}, the core technology enabling training of neural networks with millions of parameters using gradient descent. AD libraries~\citep{Paszke2017AutomaticDI,RevelsLubinPapamarkou2016,DBLP:journals/corr/abs-1907-07587} make it possible to take exact derivatives of arbitrary computable functions by using the chain rule to compose the gradients of individual operations.

The approach of creating automatically-differentiable mechanistic models that can be combined with deep neural networks is known as \emph{differentiable programming}~\citep{DBLP:journals/corr/abs-1907-07587}. Differentiable programming has recently been applied to engineering problems, with demonstrated benefits for challenging optimization problems in various domains. Constructing a differentiable ray-tracer~\citep{Li:2018:DMC} and image-processing algorithms~\citep{Li:2018:DPI,Sitzmann:2018:EndToEndCam} simplifies the inverse problem of fitting parameters describing the lighting, materials and objects in a scene from a photograph, as well as the forward problem of optimizing image-processing pipelines. Differentiable rigid-body physics engines make it possible to train deep learning controllers for robots in accurate environments~\citep{DBLP:journals/corr/DegraveHDW16}. These simulators can also be combined with neural networks to model physical systems based on video and predict their future behavior~\citep{NIPS2018_7948}.

Recently, deep learning methods have been brought to bear on strong lensing analyses. \citet{Hezaveh:2017sht} and \citet{PerreaultLevasseur:2017ltk} applied a convolutional neural network (CNN)~\citep{CNN} to infer the parameters of singular isothermal ellipsoid lenses~\citep{1998ApJ...495..157K,1994A&A...284..285K} with unprecedented speed. This CNN was later coupled to an optimizer controlled by another CNN that performs a linear inversion to recover the source's pixelated brightness profile without requiring regularization hyperparameters~\citep{Morningstar:2019szx}. Both CNNs were trained using supervised learning on mock lensing datasets. Very recently, supervised CNNs have been trained using toy models of lensing systems containing substructure to differentiate between different DM models~\citep{Alexander:2019puy} and infer parameters in the subhalo mass function~\citep{Brehmer:2019jyt}. Moreover, in \citet{DiazRivero:2019hxf} a supervised CNN  trained to classify whether or not a lensing system observation contained detectable substructure, thus identifying images worthy of follow-up observations and analyses.

In this work we construct the first differentiable programming-based strong lensing analysis pipeline, consisting of a deep generative model for the source galaxy brightness profile and a physics model for the lens. This approach removes the need for manual modeling or hyperparameter optimization and speeds up inference using gradient-based techniques. In particular, we use a variational autoencoder (VAE)~\citep{2013arXiv1312.6114K,2014arXiv1401.4082J} to learn a parametric description of the source galaxy's light. In contrast with the aforementioned deep learning strategies for lensing, the VAE is trained in an unsupervised manner on unlensed galaxies; a similar VAE was previously constructed in \citet{Ravanbakhsh:2016xpe}. The lensing physics is implemented by solving the Poisson equation for analytical models of the lens and external shear. Since our analysis pipeline is modularized, it is possible to change parameters' priors or include additional lensing effects such as multiple sources and lenses, and line-of-sight halos and subhalos without having to retrain the source VAE. The source model can also be improved independently from the rest of the code.

Our pipeline is implemented in the PyTorch machine learning framework~\citep{Paszke2017AutomaticDI}, which contains an automatic differentiation engine and graphical processing unit (GPU)-accelerated functions for array computations. As with the differentiable programming examples mentioned above, pervasive AD makes it straightforward to fit the high-dimensional parameter vector of our lensing model using gradient-based methods. We also exploit the Pyro~\citep{bingham2018pyro} probabilistic programming language to characterize the uncertainties of our parameter fits. Pyro is capable of sampling model parameters, computing likelihoods, and automatically performing inference with tools such as Markov Chain Monte Carlo (MCMC) and variational inference~\citep{2016arXiv160100670B,2012arXiv1206.7051H} on arbitrary probabilistic models, even those with stochastic control flow. While probabilistic programming has existed for a long time~\citep{Lunn2000,doi:10.1002/sim.3680}, it has only recently been integrated with automatic differentiation, making parameter inference possible even for models including neural networks~\citep{tran2016edward,JSSv076i01,DBLP:journals/corr/abs-1810-00873,bingham2018pyro,ge2018t,Charnock:2019rbk}. In the context of high energy physics, recent work has reframed the Sherpa event generator using probabilistic programming, enabling more efficient simulation of rare events~\citep{DBLP:journals/corr/abs-1807-07706, 2019arXiv190703382G}. In our present work, we again leverage the differentiability of our pipeline by employing Hamiltonian Monte Carlo (HMC)~\citep{2012arXiv1206.1901N,2017arXiv170102434B}, a gradient-based MCMC method, to efficiently sample from the high-dimension posterior for the lens and source parameters. The unique meshing of an exact physics model with a deep generative source model in an AD-compatible, probabilistic programming framework makes the current paper one of the first differentiable probabilistic programs to our knowledge.

We begin this paper by discussing the variational autoencoder model for source galaxies in Section~\ref{sec:vae}, and describing the training procedure and validation tests. In Section~\ref{sec:lensing} we review the physics of strong gravitational lensing, and introduce all the ingredients required to generate lensed images. In particular, we define the lens and the source models considered in the present paper. Section~\ref{sec:pipeline} delineates our lensing inference pipeline. In Section~\ref{sec:results} we test the pipeline on two mock galaxy-galaxy lensing systems, and discuss parameter fitting and posterior analysis. We draw our conclusions in Section~\ref{sec:conclusions}.

\section{Deep generative models for source galaxies}
\label{sec:vae}

This section describes how we use a variational autoencoder (VAE) to construct a parametric model for source galaxy light. After reviewing VAEs and explaining the architecture we selected, we describe how ours is trained and present tests validating the performance of the parametric model we use it to construct.

\subsection{Variational Autoencoders}

Natural images (such as those of galaxies) lie on or near a low-dimensional submanifold in the space of all possible images~\citep{BelkinThesis}. This motivates the concept of \emph{probabilistic latent variable models}~\citep{Bishop1998}, where a datum $\mathbf{x}$ (such as a galaxy image) is related to a latent variable $\mathbf{z}$ through a conditional probability density $p(\mathbf{x} | \mathbf{z})$, and the latent variables are assumed to have a prior density $p(\mathbf{z})$. We will use this to construct a parametric model for galaxy images, where $\mathbf{z}$ maps onto the mean of the decoding distribution and has prior $p(\mathbf{z})$.

\begin{figure}
    \centering
    \includegraphics[width=\linewidth]{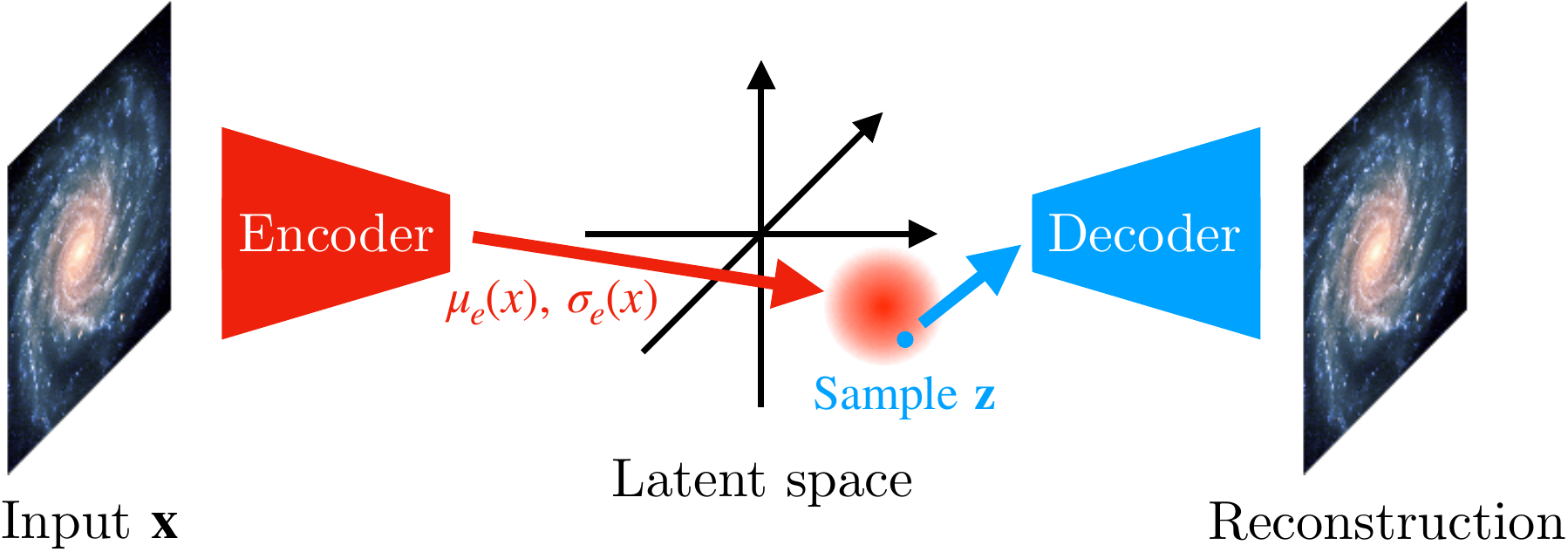}
    \caption{\textbf{Diagram of a variational autoencoder.} The diagram shows how an input image $\mathbf{x}$ is passed through the encoder (red trapezoid) to yield an encoding distribution (red blob) with mean and standard deviation $\mu_e(\mathbf{x})$ and $\sigma_e(\mathbf{x})$. A point is then sampled from this distribution (blue dot) and passed through the decoder (blue trapezoid) to yield a reconstructed image.}
    \label{fig:vae}
\end{figure}

The variational autoencoder (VAE) was constructed to efficiently approximate models of this form~\citep{2013arXiv1312.6114K,2014arXiv1401.4082J}. It approximates the conditional density $p(\mathbf{x} | \mathbf{z})$ with a decoder $d_\theta(\mathbf{x} | \mathbf{z})$ whose parameters are functions represented by a neural network; $\theta$ represents the network's parameters. The VAE also includes an encoder, $e_\phi(\mathbf{z} | \mathbf{x})$, that similarly approximates the conditional distribution $p(\mathbf{z} | \mathbf{x})$ using a neural network with parameters $\phi$. Figure~\ref{fig:vae} illustrates this structure. The decoder and encoder are typically both taken to be diagonal Gaussians:
\begin{align}
    d_\theta(\mathbf{x} | \mathbf{z}) &= \mathcal{N}(\mathbf{x} | \mu_d(\mathbf{z}), \sigma_d)\\
    e_\phi(\mathbf{z} | \mathbf{x}) &= \mathcal{N}(\mathbf{z} | \mu_e(\mathbf{x}), \sigma_e(\mathbf{x})).
\end{align}
In our notation $\mathcal{N}(x | \mu, \sigma)$ denotes that $x$ follows a normal distribution with mean $\mu$ and standard deviation $\sigma$. The functions $\mu_d$, $\mu_e$ and $\sigma_e$ are implemented using neural networks, and the decoder's standard deviation $\sigma_d$ is a constant hyperparameter. We identify $\sigma_d$ with the approximate standard deviation of the Gaussian noise in our training dataset.

Training VAE requires a dataset $\{ \mathbf{x} \}$ as well as selecting the latent space's dimensionality and prior $p(\mathbf{z})$, which is typically taken to be $\mathcal{N}(0, I)$.\footnote{Several works have studied more complex prior distributions, including learnable ones~\citep{DBLP:journals/corr/ChenKSDDSSA16,DBLP:journals/corr/DilokthanakulMG16,vamprior,DBLP:journals/corr/abs-1711-00464,DBLP:journals/corr/GoyalHLWX17}.} Ideally, the VAE would be trained by maximizing the marginal likelihood of the training dataset, which requires computing the integral
\begin{equation}
    p_\theta(\mathbf{x}) = \int d\mathbf{z}\ d_\theta(\mathbf{x} | \mathbf{z}) p(\mathbf{z})
\end{equation}
for each training point. Note that the marginal likelihood only depends on the decoder. However, this integral is generally intractable. The difficulty is circumvented by defining an alternative objective function called the evidence lower bound (ELBO), whose derivation is reviewed in Appendix~\ref{app:elbo}. The reason the encoder network was introduced is because it is required to compute the ELBO, whose value for each training point is given by\footnote{The objective eq.~\eqref{eq:elbo_objective} is equivalent to the $\beta$-VAE~\citep{betavae} objective obtained by taking $\beta = \sigma^2$ and setting the decoder's standard deviation to 1.}
\begin{align}
\label{eq:elbo_objective}
    \log p_\theta(\mathbf{x}) &\geq \sum_i \elbo(\mathbf{x}; \theta, \phi)\\
    &\equiv \mathbb{E}_{e_\phi(\mathbf{z} | \mathbf{x})} \left[ \log d_\theta(\mathbf{x} | \mathbf{z}) \right] - D_{KL}\left[ e_\phi(\mathbf{z} | \mathbf{x}) || p(\mathbf{z}) \right]. \notag
\end{align}
The first term is related to the quality of the reconstruction obtained by passing an image through the encoder followed by the decoder: maximizing this term improves reconstruction quality. The function $D_{KL}[\, \cdot \, || \, \cdot \, ]$ in the second term is the Kullback-Leibler divergence, which measures the difference between its two argument probability distributions. This term's maximization drives the averaged encoding distribution to look more like the prior $p(\mathbf{z})$. The VAE is trained to maximize the ELBO with stochastic gradient descent by taking random minibatches of training images. The second term can often be computed analytically, which makes computing its gradient straightforward. A Monte Carlo estimate of the first term can be computed. Appendix~\ref{app:reparameterization_trick} explains how the derivative of this estimate can be computed.

Our VAE architecture choice is influenced by three widespread trends in deep convolutional neural network design~\citep{dcgan}: replacing pooling functions with strided convolutions, relying on convolutional layers rather than fully-connected ones when possible, and interleaving batch normalization layers~\citep{batchnorm}. In more detail, the encoder uses five blocks made up of strided convolutions, batch normalization layers and LeakyReLU (leaky rectified linear unit) activation functions~\citep{leakyrelu}. The output from these blocks is processed by two separate dense layers, which give the $\mu_e(\mathbf{x})$ and $\sigma_e(\mathbf{x})$, the parameters of the encoding distribution. The decoder is similarly built from five blocks consisting of a transposed convolution, batch normalization layers and ReLU (rectified linear unit)~\citep{relu} activation function. The output of last block is passed through a $\tanh$ activation function to produce $\mu_d(\mathbf{x})$, the pixel values of which are thus restricted to lie within $[-1, 1]$. The standard deviation of the decoding distribution was set to $\sigma_d = 1/50$, which is the approximate standard deviation of the noise in our training dataset. We also studied the impact of making $\sigma_d$ a trainable parameter (as recommended in \cite{2019arXiv190305789D}), in which case we find it converges to approximately this value. We used 64 latent-space dimensions. Complete details of the architecture and weight initializations can be found in Appendix~\ref{app:vae_architecture}, where we also describe other architectures with which we experimented.

After training the VAE as described in the next subsection, we obtain a parametric model for galaxy images where $\mathbf{z}$ corresponds to the image $\mu_d(\mathbf{z})$, and has prior $p(\mathbf{z})$. However, it is a well-known and difficult-to-solve problem that the assumed prior $p(\mathbf{z})$ does not actually match the ``aggregate prior'' obtained by encoding the full training dataset,~\citep{2014arXiv1401.4082J,DBLP:journals/corr/abs-1711-00464}
\begin{equation}
    q_\phi(\mathbf{z}) \equiv \frac{1}{N} \sum_{i=1}^N e_\phi\left( \mathbf{z} | \mathbf{x}^{(i)} \right).
\end{equation}
This mismatch can cause problems when performing maximum a posteriori (MAP) estimation of $\mathbf{z}$ for a given galaxy image. The assumed prior can drag $\mathbf{z}$ into unrealistic regions of parameter space far from the location preferred by the likelihood, where the corresponding decoded image $\mu_d(\mathbf{z})$ does not look like a galaxy.

A variety of methods have been proposed to address this problem, including constructing simple priors using $q_\phi(\mathbf{z})$~\citep{vamprior,Otten:2019hhl}, fitting $q_\phi(\mathbf{z})$ with a second VAE after training the primary one~\citep{2019arXiv190305789D}, or using normalizing flows~\citep{2015arXiv150505770J,2016arXiv160604934K,2017arXiv170507057P,2018arXiv180305649V}. Here we follow the simpler approach of creating a weakly-informative prior for $\mathbf{z}$ by fitting a multivariate normal distribution to the set of encoded means of the training data $\{ \mu_e ( \mathbf{x}^{(i)} ) \}_{i=1}^N$ and rescaling its covariance matrix by a factor of $9$. This prior roughly confines the latent variable to realistic regions of the latent space while remaining diffuse enough that the likelihood drives MAP estimates of $\mathbf{z}$ from observations.

\subsection{Training}

We construct a dataset for training the VAE starting from 56,062 images of galaxies in the COSMOS field taken by the Hubble Space Telescope~\citep{Scoville:2006vq,Scoville:2006vr,Koekemoer:2007sr}. These were used for the GREAT3 weak gravitational lensing challenge~\citep{Mandelbaum:2013esa,great3}. An estimate of the pixel noise is included for each image, which is assumed to be Gaussian and uncorrelated. Parameters for Sersic profiles fit to each of the galaxies are also provided. Details about the image processing and parameter fits can be found in Appendix E of \citet{Mandelbaum:2013esa}.

We discard the small number of galaxies for which the Sersic fits failed or gave unphysical parameters. The dimensions of the images in the dataset vary, so we remove any that are smaller than $64\times64$ pixels. Images with unequal width and height are cropped into squares, and all are then downscaled to $64 \times 64$ pixels. Finally, many of the galaxies are extremely bright and small, and some are nearly indistinguishable from the pixel noise. In our experiments both of these degraded the quality of the VAE's reconstructions, in the former case by biasing it towards producing only compact, bright galaxies and in the later case by degrading the fidelity of the reconstructions. We find a useful, very heuristic way of removing these is to make a cut on the image ``signal-to-noise'' quantity $\max(\mathcal{I}_{1/4}) / \sigma_\u{noise}$, where $\sigma_\u{noise}$ is the standard deviation of the pixel noise and $\mathcal{I}_{1/4}$ is the image downscaled by a factor of 4. Restricting this quantity to lie between $15$ and $50$ reduces the dataset to 17,543 images. We then split these into a training, test and validation datasets consisting of 15,500, 500 and 1,543 images respectively. This is a fairly small training dataset by industrial machine learning standards~\citep{ImageNet}, but could be greatly augmented by future astronomical observations.

The training and validation sets are preprocessed by dividing each image by its maximum pixel value. The VAE's parameters are optimized to minimize the ELBO of the training using the Adam optimizer~\citep{AdamOptim}, with a learning rate of $10^{-6}$, minibatch size of $32$ and momentum parameters $(\beta_1, \beta_2) = (0.5, 0.999)$. The mean ELBO for the images in the validation set is monitored during training. While this decreases at first, it inevitably increases as the VAE starts overfitting the training data. We terminate training at this point ($\sim450$ epochs in practice). The test-set images are not seen by the VAE during training.

\subsection{Validation tests}

\begin{figure}
    \centering
    \includegraphics[width=\linewidth]{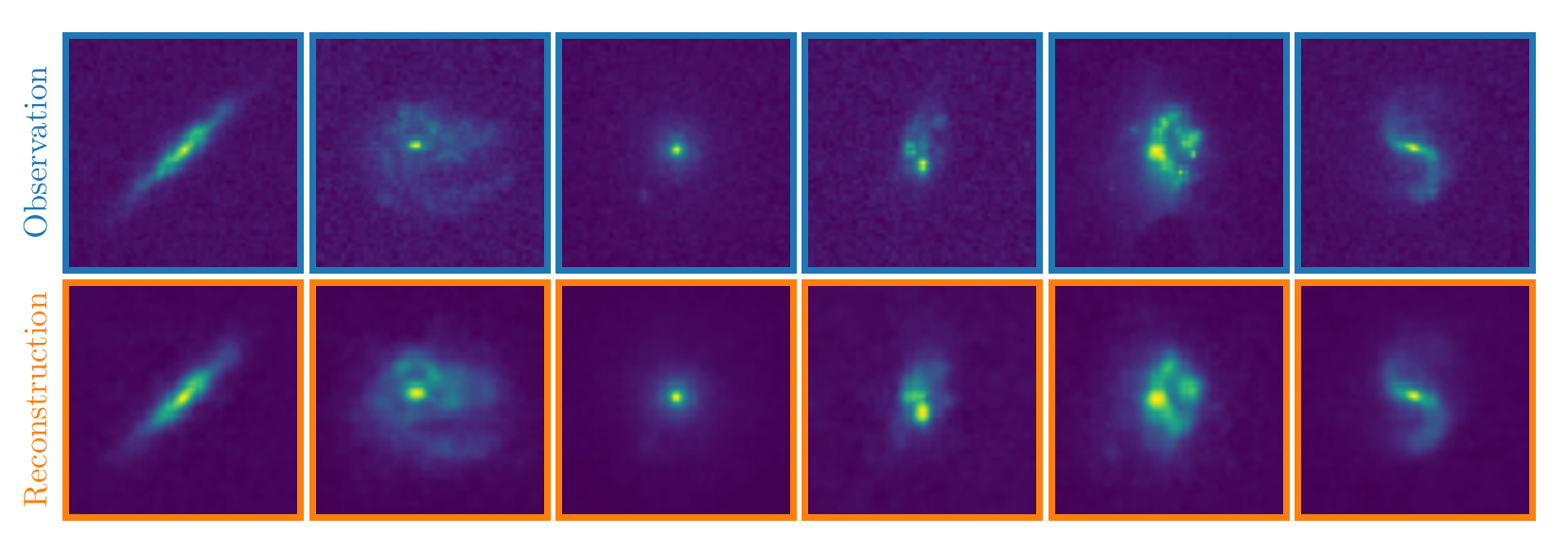}
    \caption{\textbf{Reconstructions of galaxies from the test set.} The input images are shown in the top row and the reconstructions in the bottom row. The reconstructions were obtained by passing the observations $\mathbf{x}$ through the encoder, sampling $\mathbf{z} \sim \mathcal{N}(\mu_e(\mathbf{x}), \sigma_e(\mathbf{x}))$ from the encoded distribution, and taking the mean of the decoded distribution $\hat{\mathbf{x}} = \mu_d(\mathbf{z})$.}
    \label{fig:reconstructions}
\end{figure}

To assess the quality of our parametric model $\mu_d(\mathbf{z})$ for galaxy images, we present the reconstructions of galaxies from the test set that the model has not seen during training in Fig.~\ref{fig:reconstructions}. We can observe that the reconstructions of our VAE model are denoised versions of the original images that reliably contain galaxy substructure. This indicates the latent space learned by the VAE contains a point corresponding to each of these galaxies, even though they have complex and varied morphologies.

\begin{figure}
    \centering
    \includegraphics[width=\linewidth]{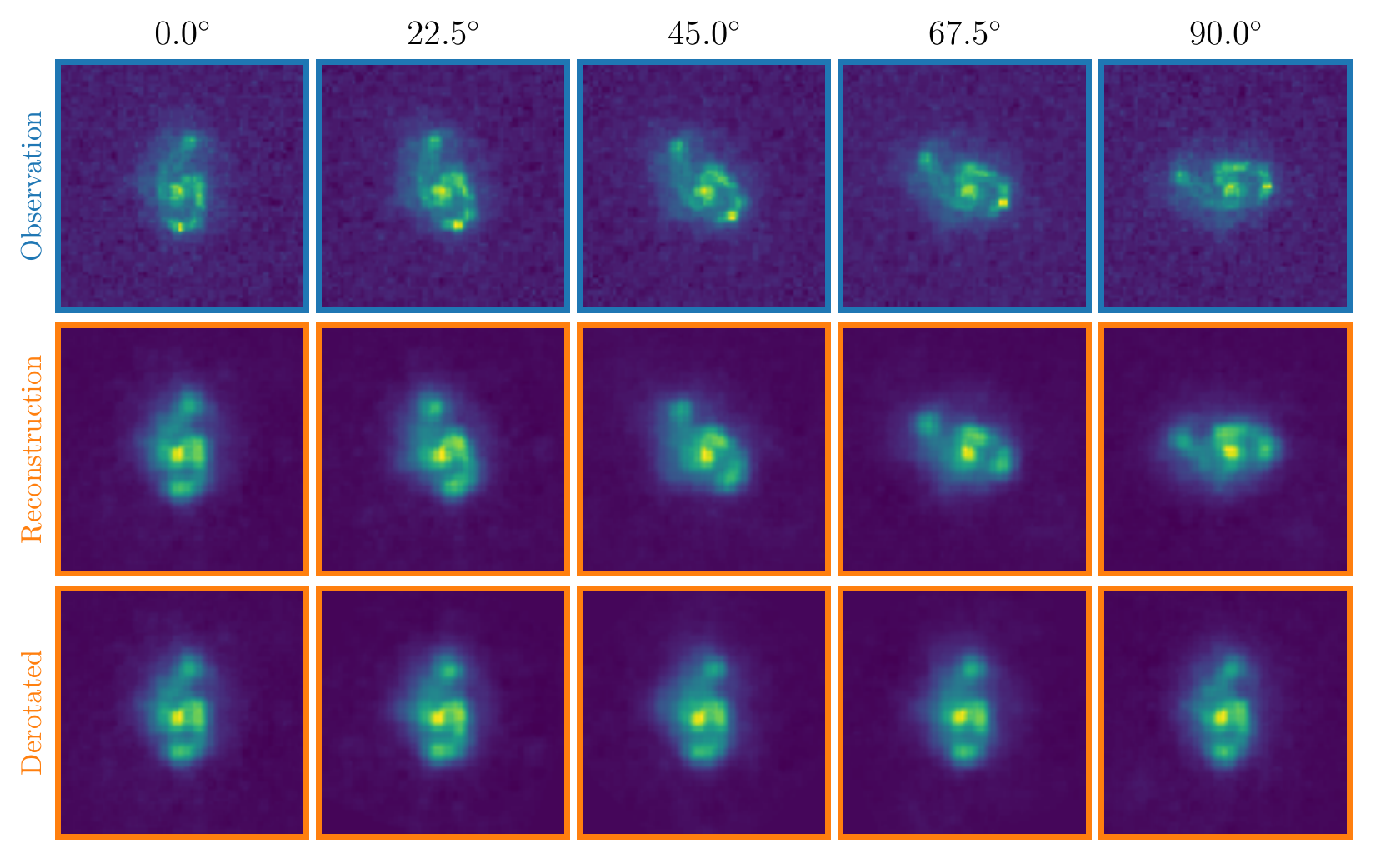}
    \caption{\textbf{Approximate rotational invariance of reconstructions.} The first row shows a galaxy image from the test set and versions rotated counterclockwise by the amount indicated above each column, with new pixels filled using the known noise distribution. The second row shows the corresponding VAE reconstructions, with derotated versions shown in the third row to make comparison simpler. The same color scale is used in each subplot.}
    \label{fig:rotated_reconstructions}
\end{figure}

It is also important that the VAE's reconstructions are equivariant under transformations such as rotations, since the parameter inferences by the analysis pipeline should be as well. In Fig.~\ref{fig:rotated_reconstructions}, a galaxy image from the test set is rotated by various angles. Each of the rotated images is then derotated for comparison. The reconstructed images are once again denoised versions of the input images, and it can be seen by inspection that the derotated reconstructions are nearly identical. From this we conclude that our generative model has (approximately) learned rotational equivariance, even though the training dataset was not augmented with rotated images to teach this explicitly.

\section{Strong lensing}
\label{sec:lensing}

Here we review the physics of strong gravitational lensing, describe how we model the main lens and external shear, and explain how we construct the source model using the variational autoencoder from the previous section.

\subsection{Strong lensing physics}

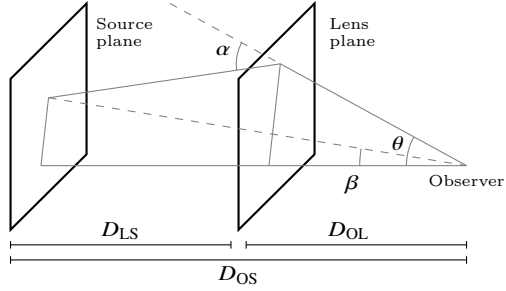
\begin{figure}
    \centering
    \begin{tikzpicture}
\draw[black,thick] (0, 0) -- (1, 1) -- (1, 3) -- (0, 2) -- cycle;
\draw[black,thick] (3, 0) -- (4, 1) -- (4, 3) -- (3, 2) -- cycle;
\draw[gray, thin] (0.4, 0.85) -- (6, 0.85);
\draw[gray, thin] (0.4, 0.85) -- (0.5, 1.75); 
\draw[gray, thin, dashed] (0.5, 1.75) -- (6, 0.85);
\draw[gray, thin] (3.4, 0.85) -- (3.55, 2.2); 
\draw[gray, thin] (0.5, 1.75) -- (3.55, 2.2) -- (6, 0.85);
\draw[gray, thin, dashed] (2.098, 3) -- (3.55, 2.2); 
\node[] at (6, 0.65) {\scriptsize Observer};
\node[align=left] at (4.5, 2.6) {\scriptsize Lens\\[-1ex] \scriptsize plane};
\node[align=left] at (1.5, 2.6) {\scriptsize Source\\[-1ex] \scriptsize plane};
\node[] at (2.8, 2.375) {\footnotesize $\alpha$};
\begin{scope}
    \path[clip] (2.098, 3) -- (3.55, 2.2) -- (0.5, 1.75) -- cycle;
    \node[circle,draw=gray,minimum size=33pt] at (3.55, 2.2) (circ) {};
\end{scope}
\node[] at (5.1, 1.15) {\footnotesize $\theta$};
\begin{scope}
    \path[clip] (6, 0.85) -- (3.4, 0.85) -- (3.55, 2.2) -- cycle;
    \node[circle,draw=gray,minimum size=45pt] at (6, 0.85) (circ) {};
\end{scope}
\node[] at (4.5, 0.6) {\footnotesize $\beta$};
\begin{scope}
    \path[clip] (6, 0.85) -- (0.4, 0.85) -- (0.5, 1.75) -- cycle;
    \node[circle,draw=gray,minimum size=80pt] at (6, 0.85) (circ) {};
\end{scope}
\draw[black, thin] (0, -0.2) -- (2.9, -0.2);
\draw[black, thin] (0, -0.25) -- (0, -0.15);
\draw[black, thin] (2.9, -0.25) -- (2.9, -0.15);
\node[] at (1.45, -0.015) {\footnotesize $D_{\mathrm{LS}}$};
\draw[black, thin] (3.1, -0.2) -- (6, -0.2);
\draw[black, thin] (3.1, -0.25) -- (3.1, -0.15);
\draw[black, thin] (6, -0.25) -- (6, -0.15);
\node[] at (4.45, -0.015) {\footnotesize $D_{\mathrm{OL}}$};
\draw[black, thin] (0, -0.4) -- (6, -0.4);
\draw[black, thin] (0, -0.45) -- (0, -0.35);
\draw[black, thin] (6, -0.45) -- (6, -0.35);
\node[] at (3, -0.6) {\footnotesize $D_{\mathrm{OS}}$};
\end{tikzpicture}
    \caption{\textbf{Diagram of a gravitational lensing system.} The simplest galaxy-galaxy lensing system is represented by a background and a foreground galaxy, which define the source (S) and the lens plane (L), respectively. The quantities $D$ are the angular diameter distances between the different planes and the observer (O). The two-dimensional angular coordinates, $\bm{\beta}$ and $\bm{\theta}$, are related through the displacement field $\bm{\alpha}$ by the lens equation~\eqref{eq:lens_eq}.}
    \label{fig:lensing_diagram}
\end{figure}

A gravitational galaxy-galaxy lensing system mainly consists of a background galaxy, playing the role of the source, and a foreground galaxy, acting as the main lens that deflects the light through its gravitational potential. In the thin lens approximation, the relation between the two-dimensional angular coordinates of the lens plane $\bm{\theta}$ and the ones in the source plane $\bm{\beta}$ is encoded by the lens equation~\citep{1994A&A...284..285K,Treu:2010uj}
\begin{equation}
    \bm{\beta} = \bm{\theta} - \bm{\alpha}(\bm{\theta}) \,,
    \label{eq:lens_eq}
\end{equation}
where $\bm{\alpha}$ is the displacement field, which defines the deflection experienced by the light ray. The geometry of the system is displayed in Fig.~\ref{fig:lensing_diagram}. The displacement field depends on the Newtonian gravitational potential related to the mass distribution of the foreground galaxy. By means of the Poisson equation, it can be expressed as
\begin{equation}
    \bm{\alpha} = \frac{4\,G_N}{c^2}\frac{D_{\rm OL}D_{\rm LS}}{D_{\rm OS}}\int \Sigma(\bm{\theta}') \frac{\bm{\theta} - \bm{\theta}'}{|\bm{\theta} - \bm{\theta}'|^2}d^2\theta'\,,
    \label{eq:deflection}
\end{equation}
where $\Sigma$ is the surface mass density of the lens, and the quantities $D_{\rm OL}$, $D_{\rm OS}$, $D_{\rm LS}$ are the angular diameter distances between the observer and the lens, the observer and the source, the lens and the source, respectively. Moreover, $G_N$ is the gravitational constant while $c$ is the speed of light.

Since the lens equation (eq.~\eqref{eq:lens_eq}) preserves the surface brightness (photon flux density per unit angular area), the image of the system in the lens plane, denoted as $\mathcal{I}_{\rm lens}$, is simply obtained by evaluating the source light distribution $\mathcal{I}_{\rm src}$ on the lens plane. Hence, we have\footnote{Note that we do not consider the light distribution of the foreground galaxy since it is in general subtracted in real data analyses.}
\begin{equation}
    \mathcal{I}_{\rm lens} (\bm{\theta}) = \mathcal{I}_{\rm src}(\bm{\theta} - \bm{\alpha}(\bm{\theta})) \,.
    \label{eq:lens_bright}
\end{equation}
This equation is solved on a square pixel grid defined in the lens plane according to the observed image. In particular, we consider a grid of $256\times256$ pixels with angular size $10\us{arcsec}\times 10\us{arcsec}$, corresponding to a pixel size of $~0.04~{\rm arcsec}$.\footnote{For reference, this is the pixel size of the optical/UV CCDs of WFC3 Hubble~\citep{2012wfci.book.....D}.} Then, the predicted lensed image $\mathcal{I}_{\rm pred}$ is obtained by taking into account the Point Spread Function (PSF), which defines how a point-like source (a pixel) is spread due to atmospheric distortions and defects in the optics, and the noise from instrumental and astrophysical backgrounds. In the present paper, we consider a symmetric two-dimensional Gaussian PSF with a standard deviation of 0.05~arcsec. Moreover, to each pixel we add uncorrelated Gaussian noise with $\sigma_\u{noise} = 0.333$, $\sigma_\u{noise} = 0.1$ and $\sigma_\u{noise} = 0.0333$ for mock images with low, medium and high signal-to-noise ($S/N$) ratio, respectively. Hence, we have
\begin{equation}
    \mathcal{I}_{\rm pred} = \mathcal{N}\left({\rm PSF} * \mathcal{I}_{\rm lens}, \sigma_{\rm noise}\right) \,,
\end{equation}
where the symbol $*$ stands for the mathematical convolution between the Point Spread Function and the image in the lens plane.

The two main ingredients describing a gravitational lensing system are, therefore, the total mass distribution of the lens $\Sigma$ (the lens model) and the surface brightness profile of the background source $\mathcal{I}_{\rm src}$ (the source model).

\subsection{Lens model \label{sec:lens_model}}

In a typical lensing system, the dominant contributions to the total displacement field $\bm{\alpha}$ come from the smooth main halo (mh) of the foreground galaxy and the external shear (ext), namely 
\begin{equation}
    \bm{\alpha} = \bm{\alpha}_{\rm mh} + \bm{\alpha}_{\rm ext}.
    \label{eq:tot_isplacement}
\end{equation}
For the first component, we consider the so-called Singular Power-Law Ellipsoid (SPLE) profile to model the surface mass distribution of the main halo. Such profiles can fit gravitational potentials of lenses in images of galaxy-scale strong lensing systems at the percent level (see e.g.~\citet{Suyu:2008zp}). In this case, we have~\citep{Kassiola1993,Barkana:1998qu}
\begin{equation}
    \Sigma_{\rm mh} = \Sigma_{\rm cr}\frac{3-\gamma}{2}\left[\frac{\rho\left(\bm{\theta}', q\right)}{r_{\rm Ein}}\right]^{1-\gamma} \,,
    \label{eq:sple}
\end{equation}
where $\Sigma_{\rm cr} = c^2 D_{\rm OS} / (4 \, \pi\,G\,D_{\rm OL}\,D_{\rm LS})$ is the critical surface mass density, $\gamma$ is the slope, $r_{\rm Ein}$ denotes the Einstein radius, and $\rho$ encodes the dependence on the position. In the special case $\gamma = 2$ this distribution reduces to the one for a singular isothermal ellipsoid~\citep{1994A&A...284..285K}. In the coordinates system $\bm{\theta}'(\bm{\theta}-\bm{\theta}_{\rm lens}, \phi)$ with origin in the centroid of the foreground galaxy (denoted as $\bm{\theta}_{\rm lens}$) and axes aligned to the minor and major axes of the elliptical galaxy (after a rotation of an angle $\phi$), we have
\begin{equation}
    \rho(\bm{\theta}', q) = \theta'^2_1 + \theta'^2_2/q^2 \,,
\end{equation}
with $q$ being the minor to major axis ratio of the elliptical contours of equal surface mass density. Hence, the displacement field $\bm{\alpha}_{\rm mh}$ induced by the main halo mass distribution is simply given by plugging eq.~\eqref{eq:sple} into eq.~\eqref{eq:deflection}.

Since numerical integration is difficult to implement in an automatically-differentiable manner, we instead compute $\bm{\alpha}_{\rm mh}$ by interpolating over a precomputed grid, as described in detail in Appendix~\ref{app:main_halo}. This enables computation of gradients of $\bm{\alpha}_{\rm mh}$ since the interpolation function is itself automatically differentiable.

The external shear contribution represents the additional angular structure provided by additional matter distribution in the cluster where the galaxy is located. The corresponding displacement field is parametrized as
\begin{equation}
    \bm{\alpha}_{\rm ext} = \left(\begin{array}{cc}\gamma_1 & \gamma_2 \\ \gamma_2 & -\gamma_1\end{array}\right)\bm{\theta}.
    \label{eq:rho}
\end{equation}
Our lens model is thus completely defined by a set of 8 parameters: $\mathbf{\Theta}_\u{lens} \equiv \{ \gamma_1, \gamma_2, \phi, q, \gamma, r_{\rm Ein}, \theta_{\rm lens, 1}, \theta_{\rm lens,2} \}$.

\subsection{Source model \label{sec:source_model}}

The VAE's decoder $d_\theta(\mathbf{z})$ provides a parametrized model for $64\times64$-pixel galaxy images, which is the basis for our source model. One of its parameters is the 64-dimensional vector $\mathbf{z}$ specifying a point in the VAE's latent space. As described earlier, the prior for $\mathbf{z}$ is defined by fitting a multivariate normal distribution to the means of the encoded distributions of the training points and increasing its covariance by a factor of $9$. We introduce four other parameters to complete the model: $\theta_{\u{src}, 1}$, $\theta_{\u{src}, 2}$, $s$ and $\iota$. The first two are the position of the center of the decoded image $d(\mathbf{z})$. The second specifies the spatial scale of the image and the third is the normalization of the pixel intensities. In this work we adopt the priors
\begin{align}
    \label{eq:src_priors_start}
    \theta_{\u{src},1}, \theta_{\u{src},2} &\sim \mathcal{N}(0, 0.1)\\
    s &\sim \mathcal{N}(5, 1)\\
    \iota &\sim \mathcal{N}(1, 0.5)
    \label{eq:src_priors_end}
\end{align}
for our mock data generation and analysis. Our overall model for the surface brightness of the source galaxy at a position $\bm{\theta} = (\theta_1, \theta_2)$ is thus
\begin{equation}
    \mathcal{I}_\u{src}(\bm{\theta} | \mathbf{\Theta}_\u{src}) = \iota \, d(\mathbf{z})\left( \frac{\theta_1 - \theta_{\u{src},1}}{s}, \frac{\theta_2 - \theta_{\u{src}},2}{s} \right),
    \label{eq:src_bright}
\end{equation}
where $\mathbf{\Theta}_\u{src} \equiv \{ \mathbf{z}, \theta_{\u{src}, 1}, \theta_{\u{src}, 2}, s, \iota \}$. We use bilinear interpolation to allow the right-hand side of this expression to be evaluated between adjacent pixels in the $64\times64$ output image from the decoder.

\section{The lensing pipeline}
\label{sec:pipeline}

\begin{figure}
     \centering
     \input{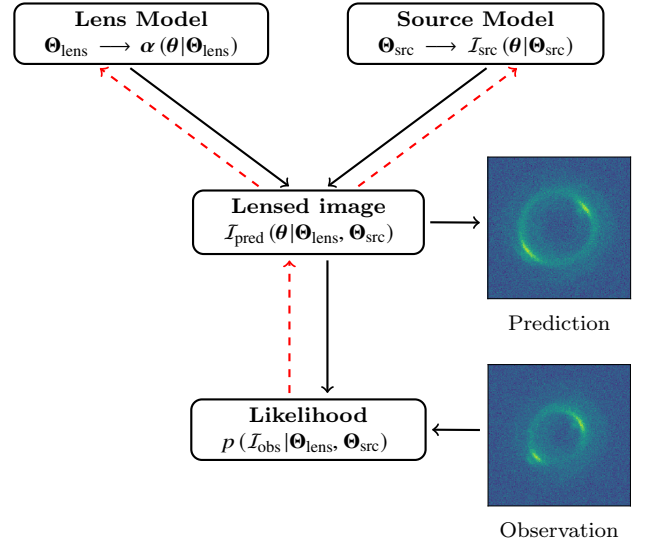}
     \caption{\textbf{Lensing Pipeline.} The forward flow (black solid lines) estimates the likelihood by comparing the observation with the predicted lensed image. This is obtained through the Lens Model and the Source Model. The backward flow (red dashed lines) computes the derivatives of the likelihood with respect to all the parameters of the models, $\mathbf{\Theta}_{\rm lens}$ and $\mathbf{\Theta}_{\rm src}$. Each box represents an independent module.}
     \label{fig:pipeline}
\end{figure}

Our lensing pipeline is unique since it combines the VAE-based source and physical lens models detailed in the previous section in a fully-differentiable manner. As sketched in Fig.~\ref{fig:pipeline}, the pipeline consists of a \emph{forward} flow (black solid lines) and a \emph{backward} one (red dashed lines).

In the forward flow, the predicted lensed image $\mathcal{I}_{\rm pred}\left(\mathbf{\Theta}_{\rm lens}, \mathbf{\Theta}_{\rm src}\right)$ is obtained once the displacement field $\bm{\alpha}\left(\bm{\theta} | \mathbf{\Theta}_{\rm lens} \right)$ and the source surface brightness $\mathcal{I}_{\rm src}\left(\bm{\theta} | \mathbf{\Theta}_{\rm src} \right)$ have been computed in the Lens Model (see Section~\ref{sec:lens_model}) and the Source Model (see Section~\ref{sec:source_model}), respectively. This image is then compared with the observed one to estimate the likelihood function given the parameters of the lens and the source models. Thanks to the differentiable programming framework, it is then possible to compute the derivatives of the likelihood function with respect to all the parameters, $\mathbf{\Theta}_{\rm lens}$ and $\mathbf{\Theta}_{\rm src}$. This step represents the backward flow of the whole pipeline.

The pipeline is implemented in a differentiable probabilistic programming framework comprised of the PyTorch~\citep{Paszke2017AutomaticDI} machine learning library and Pyro~\citep{bingham2018pyro} probabilistic programming library. PyTorch provides an automatic differentiation engine, enabling the backward flow of the pipeline. The likelihood calculations in the forward flow are made straightforward by Pyro. The VAE Source Model is constructed from neural network layers contained in PyTorch and trained using the variational inference module in Pyro. We employ Pyro's variational inference and Hamiltonian Monte Carlo modules for parameter fitting and posterior sampling our mock data analysis in the following section.

We stress that automatic differentiability is automatically guaranteed if all the pipeline is written in the differentiable programming language. Moreover, we note that the lensing pipeline is implemented so that the lens and the source models are independent building blocks. In the present paper, the former is fully based on physical models while the latter is provided by the VAE's decoder. However, thanks to the modularity of the pipeline, both can be easily modified or substituted, as can any of the priors on the lensing parameters. This fundamental feature allows one to generalize the present lensing pipeline to analyze more realistic systems and to include the gravitational effect of dark matter substructures in the lensing physics. This is left for future investigation.

\section{Results}
\label{sec:results}

In this section, we test the lensing pipeline on mock lensing system observations. We describe the generation of mock data, and discuss the results obtained by the parameter fitting and the posterior analysis for two different mock lensing systems.
\begin{figure*}
    \centering
    \includegraphics[width=\linewidth]{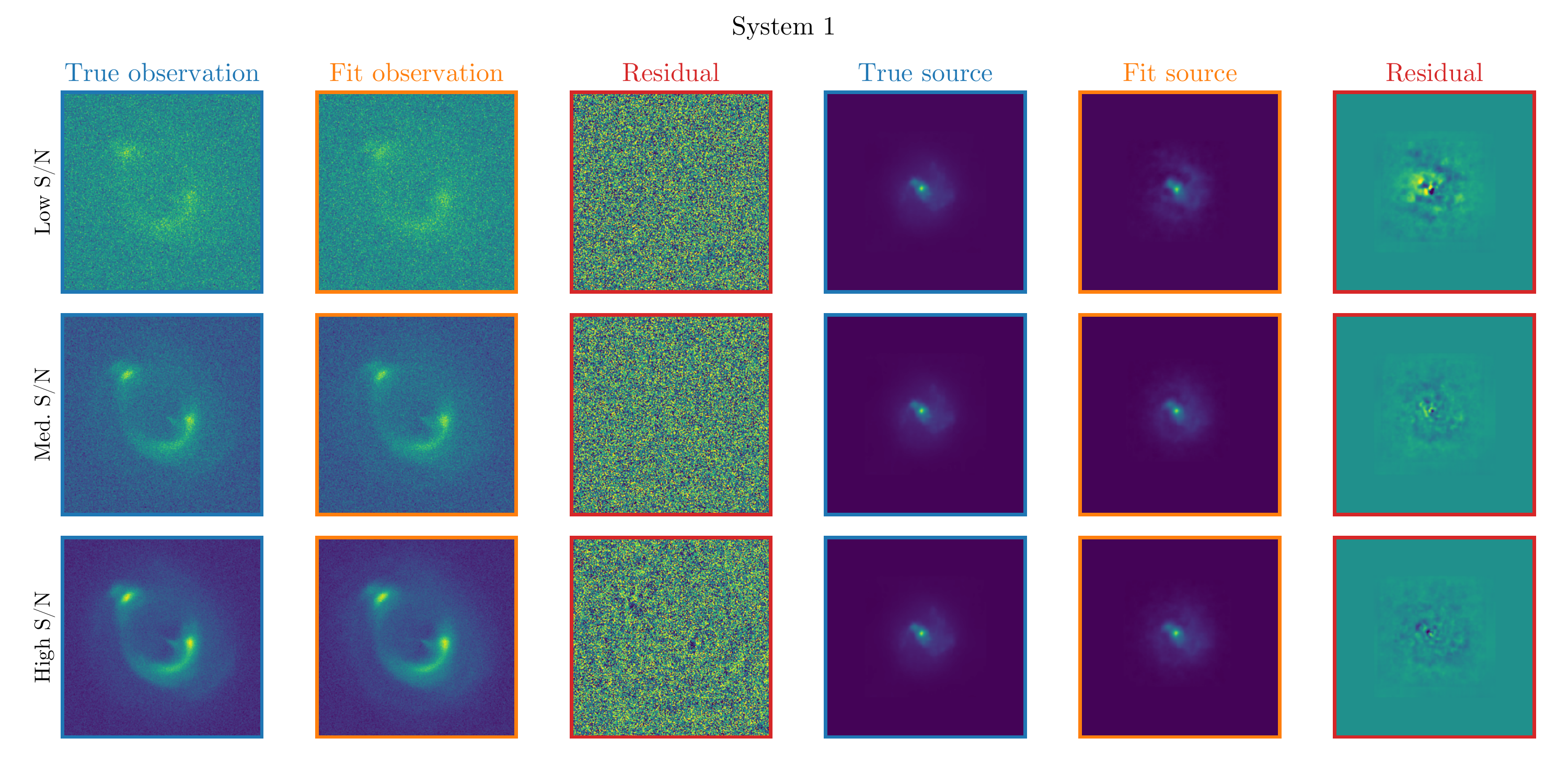}
    \caption{\textbf{Results of MAP fit of mock lensing system 1.} The different columns show the true and fit observations, the residual between these, the true and fit sources and the residuals between those. The rows correspond to different observed signal-to-noise values. Within each row, the color scales for the true and fit observations are the same, as are the scales for the true and fit sources. The observation residuals are normalized by dividing by the standard deviation of the observation noise. The source residuals are normalized by dividing by 10\% of the maximum value of the source. The color scale ranges from dark blue to bright yellow.}
    \label{fig:1_recon}
\end{figure*}
\begin{figure*}
    \centering
    \includegraphics[width=\linewidth]{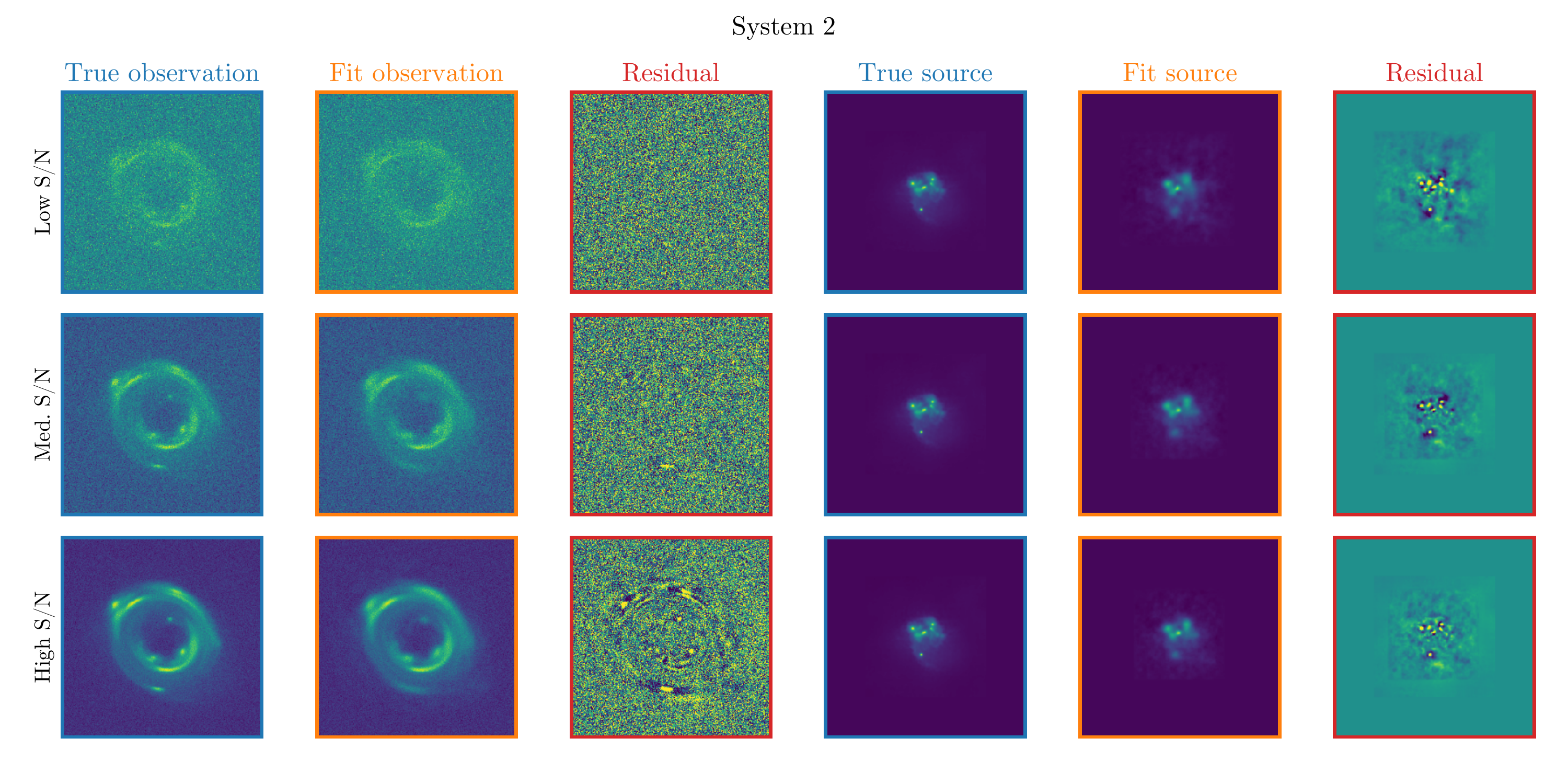}
    \caption{\textbf{Results of MAP fit of mock lensing system 2.} The subplots and scales are explained in the caption of Fig.~\ref{fig:1_recon}.}
    \label{fig:2_recon}
\end{figure*}

\subsection{Mock data}

To test our pipeline, we generate mock lensing system observations. We create mock sources by first denoising images from the test dataset using the non-local means algorithm~\citep{Buades:2005:NAI:1068508.1069066,ipol.2011.bcm_nlm}. These images are then rescaled to fill roughly the central third of the source image plane, which ensures the lensed image pixel intensities drop to zero along the image boundaries. The mock lensing parameters are determined randomly by drawing from the following distributions:
\begin{align}
    \label{eq:lens_priors_start}
    \gamma_1, \gamma_2 &\sim \mathcal{N}(0, 1)\\
    \phi &\sim \mathcal{N}(0, 1\us{rad})\\
    q &\sim \mathcal{N}(0.5, 0.5)\\
    r_\u{Ein} &\sim \mathcal{N}(1.5\us{arcsec}, 0.5\us{arcsec})\\
    \gamma &\sim \mathcal{N}(2, 0.5)\\
    \theta_{\rm lens,1}, \theta_{\rm lens,2} &\sim \mathcal{N}(0, 0.5). \label{eq:lens_priors_end}
\end{align}
The mock lensed images are then produced using the formalism from the previous section on a $256\times256$ pixel grid with angular size $10\us{arcsec}\times 10\us{arcsec}$. Finally, we convolve with the PSF and add Gaussian pixel noise.

To test how data quality impacts our pipeline's performance, we fix the pixel noise level $\sigma_{\rm noise}$ to three different values to obtain observations with low, medium and high signal-to-noise ratios. This ratio is defined by creating a mask $m$ to select only the pixels belonging to the galaxy in the observed image $\mathcal{I}_{\rm obs}$ and computing the following quantity (see e.g.~\cite{2019MNRAS.487.5143O}):
\begin{align}
    S/N(\sigma_{\rm noise}) = \frac{\sum_{i,j} m_{ij} {\mathcal{I}_{\rm obs}}_{ij}}{\sigma_{\rm noise} \sqrt{\sum_{i,j} m_{ij}}}.
\end{align}
The mask is constructed by first convolving $\mathcal{I}_{\rm obs}$ with a Gaussian $G$ with standard deviation 0.16~arcsec (four pixels) and then thresholding the blurred image using the noise level:
\begin{align}
    m_{ij} = \begin{cases}
        1 & \left( G * \mathcal{I}_{\rm obs} \right)_{ij} \geq \sigma_{\rm noise}\\
        0 & \left( G * \mathcal{I}_{\rm obs} \right)_{ij} < \sigma_{\rm noise}
    \end{cases}.
\end{align}
Changing the threshold and/or width of the Gaussian $G$ does not substantially change the value of $S/N$. The pixel noise levels and corresponding $S/N$ values for each system are shown in Tab.~\ref{tab:snrs} for completeness, though their specific values are not important for our study.
\begin{table}
    \centering
    \begin{tabular}{c c c c}
         & Low $S/N$ & Med. $S/N$ & High $S/N$ \\
        \hline
        $\sigma_{\rm noise}$ & 0.333 & 0.1 & 0.0333 \\
        \hline
        System 1 & 66 & 272 & 757 \\
        System 2 & 77 & 313 & 849\\
    \end{tabular}
    \caption{\textbf{Signal-to-noise ratios for the systems considered in this work.} The second row shows the pixel noise level while the third and fourth show the $S/N$ values for the two systems.}
    \label{tab:snrs}
\end{table}

In the rest of this paper, we focus on two particular mock systems, hereafter referred to as system 1 and system 2. Source 1 has a fairly simple spiral morphology. Source 2 has a significant amount of substructure, making it representative of complex, high-redshift source galaxies.

\subsection{Parameter fitting}

We first test our pipeline by finding the best-fit source and lens parameter values. For a given image $\mathcal{I}_\u{obs}$ we compute the maximum a posteriori parameter (MAP) estimates:
\begin{align}
    \label{eq:map}
    \hat{\mathbf{\Theta}}_\u{lens}, \hat{\mathbf{\Theta}}_\u{src} &= \max_{\mathbf{\Theta}_\u{lens},\mathbf{\Theta}_\u{src}} p(\mathbf{\Theta}_\u{lens},\mathbf{\Theta}_\u{src} | \mathcal{I}_\u{obs}),\\
    p(\mathbf{\Theta}_\u{lens},\mathbf{\Theta}_\u{src} | \mathcal{I}_\u{obs}) &\propto  p(\mathcal{I}_\u{obs} | \mathbf{\Theta}_\u{lens},\mathbf{\Theta}_\u{src}) \, p(\mathbf{\Theta}_\u{lens}) \, p(\mathbf{\Theta}_\u{src}).\notag
\end{align}
The first term on the right-hand side of the posterior is the likelihood of the observed image for fixed source and lens parameters, which is Gaussian due to our noise assumption:
\begin{equation}
    p(\mathcal{I}_\u{obs} | \mathbf{\Theta}_\u{lens},\mathbf{\Theta}_\u{src}) = \mathcal{N}(\mathcal{I}_\u{obs} | \mathcal{I}_\u{pred}(\mathbf{\Theta}_\u{lens},\mathbf{\Theta}_\u{src}), \sigma_\u{obs}). \notag
\end{equation}
The second term is the prior on the lens parameters, for which we adopt eqs.~\eqref{eq:lens_priors_start}-\eqref{eq:lens_priors_end}. The source priors are specified in eqs.~\eqref{eq:src_priors_start}-\eqref{eq:src_priors_end}, and the prior on $\mathbf{z}$ is described in Section~\ref{sec:vae}.

Since it is possible to differentiate through the full lensing pipeline, we obtain the best-fit parameters using the gradient-based Adam optimizer~\citep{AdamOptim}. Adam has been empirically shown to outperform other gradient-based methods in nonconvex optimization problems with large numbers of parameters. These fits converge after $\sim 10^4$ iterations, which takes approximately 20 (7.5) minutes on CPU (GPU).

The MAP reconstructions of the mock lensing systems 1 and 2 are shown in Figs.~\ref{fig:1_recon} and \ref{fig:2_recon}. We report the true and fit images, and the residuals between these, in the lens and in the source planes, for three values of signal-to-noise ratio. In case of system 1, the residuals in the lens plane are at the noise level even for the smallest pixel noise level (highest $S/N$) considered. This is related to the fact that the reconstruction of the source improves as the signal-to-noise ratio increases, as can clearly seen in the last column in Fig.~\ref{fig:1_recon}. This does not occur for the complex source galaxy of system~2. Indeed, as shown in Fig.~\ref{fig:2_recon}, even for the highest $S/N$ value, the VAE's decoder is not able to reproduce all the substructures exhibited in the source surface brightness. This directly affects the image reconstruction in the lens plane, and the corresponding residuals (third column in Fig.~\ref{fig:2_recon}) increase and show more structure as the signal-to-noise ratio increases.

\subsection{Posterior analysis}

To study our pipeline's parameter inference capabilities, we ran Hamiltonian Monte Carlo~\citep{DUANE1987216,2012arXiv1206.1901N,2017arXiv170102434B} to sample from the parameters' posteriors. Hamiltonian Monte Carlo (HMC) is a Markov-chain Monte Carlo (MCMC) procedure that uses Hamiltonian dynamics based on the gradient of the posterior to efficiently traverse parameter space. HMC can take larger steps than other MCMC procedures such as Metropolis-Hastings while keeping the acceptance probabilities high, leading to more efficient exploration of parameter space. After 50 steps during which the internal HMC parameters are calibrated, we use it to sample 500 times from the posterior starting from the MAP parameter estimates. This takes about 10 (6.5) hours on a CPU (GPU).
\begin{figure}
    \centering
    \includegraphics[width=\linewidth]{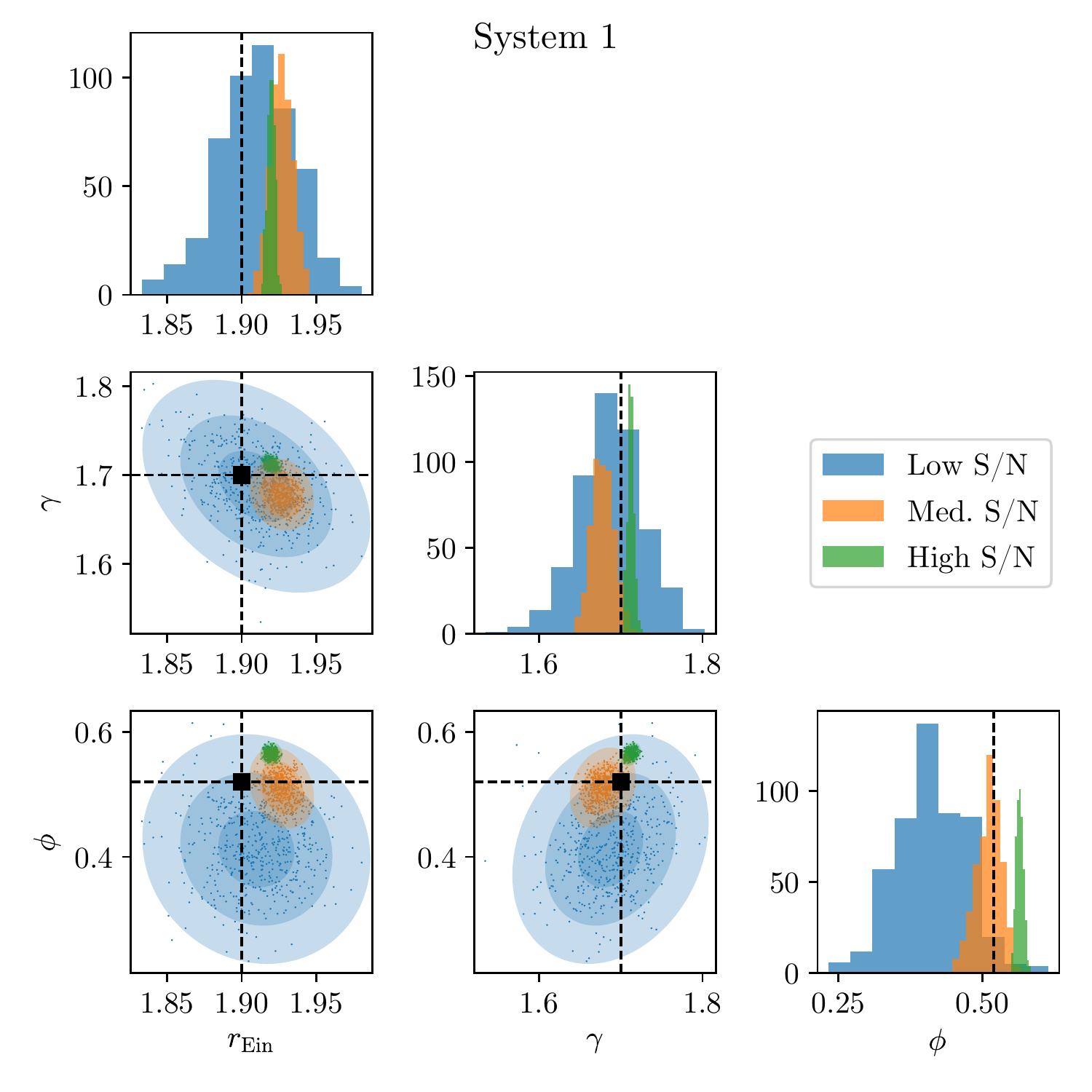}
    \caption{\textbf{Results of HMC parameter posteriors of mock lensing system 1.} The different panels show the marginalized one-dimensional and two-dimensional posteriors for a subset of lens parameter, $(r_\u{Ein}, \gamma, \phi)$. The different colors (blue, orange, green) refer to the different signal-to-noise ratios while the shading in the ellipses corresponds to 68\%, 95\% and 99\% confidence levels. The sampled points are also plotted. The black squares and dashed lines represent the true values of parameters.}
    \label{fig:1_posteriors}
\end{figure}
\begin{figure}
    \centering
    \includegraphics[width=\linewidth]{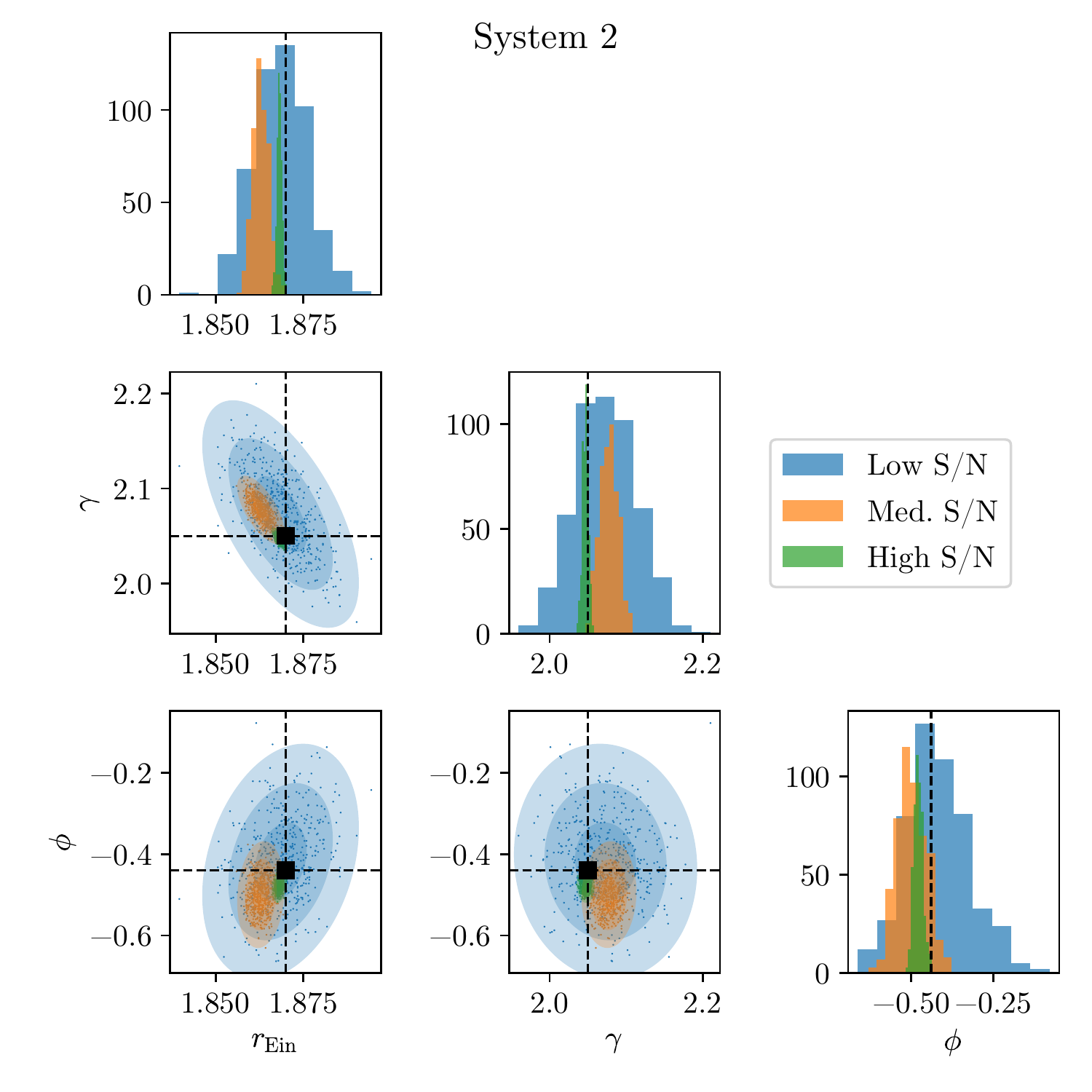}
    \caption{\textbf{Results of HMC parameter posteriors of mock lensing system 2.} The panels are explained in the caption of Fig.~\ref{fig:1_posteriors}. Note the substantial overlap between the green ellipses and black squares.}
    \label{fig:2_posteriors}
\end{figure}

The HMC results for the two systems are reported in Figs.~\ref{fig:1_posteriors} and~\ref{fig:2_posteriors}, which show the marginalized posteriors of the lens parameters $(r_\u{Ein}, \gamma, \phi)$. As expected, for both systems the posteriors shrink as the $S/N$ ratio increases. In particular, the statistical error on the parameter estimates moves from $\sim 3\%$ to $\lesssim 1 \%$ for the lowest and highest $S/N$ ratio, respectively. However, especially for the system 1, we note that our lensing pipeline gives biased estimates typically at the level of $1\%$. We hypothesize this is because the VAE generally produces slightly blurred and hazy reconstructions (most clearly visible in the last row of residual plots in Fig.~\ref{fig:2_recon}). Improving the fidelity of reconstructions and samples from VAEs is an active area of research in machine learning~\citep{DBLP:journals/corr/ZhaoSE17a,rezende2018taming,2019arXiv190305789D}. In addition to improving the VAE architecture and training procedure, our source model would benefit from a larger, higher-resolution training dataset, as will be made available by future astronomical surveys. Surprisingly, the level of bias is smaller for the system 2, even though the source-plane residuals are larger due to the presence of fine substructure.

\section{Conclusions}
\label{sec:conclusions}

We have presented the first step towards a new inference pipeline to analyze present and future strong gravitational lensing systems. The main novelty of our approach is the use of a differentiable probabilistic programming framework in which all operations are automatically differentiable with respect to the input parameters. This powerful approach makes Bayesian inference feasible for complex models with hundreds (or thousands) of free parameters thanks to efficient sampling techniques utilizing gradient descent.

Our lensing pipeline, shown in Fig.~\ref{fig:pipeline}, consists of two independent blocks describing the surface brightness of the source galaxy (Source Model) and the mass distribution of the lens galaxy (Lens Model). They are combined to generate the lensed image, which is used to estimate the likelihood and perform Bayesian inference. The advantages of this strategy are:
\begin{itemize}
    \item Exact gradients of the pipeline's output can be computed with respect to its inputs using automatic differentiation. This makes it possible to use efficient gradient-based fitting and posterior sampling procedures.
    \item Using a differentiable probabilistic programming framework allows us to integrate the variational autoencoder source model learned from unlensed galaxy images directly with a physical lensing model. Learning the source model directly from data removes the need for tuning hyperparameters to regularize the source model.
    \item We fully automatize the inference step using probabilistic programming. The lens and source parameters are fit and sampled simultaneously and can have arbitrary priors.
    \item By implementing our pipeline in the PyTorch framework, we automatically gain the ability to perform computations on graphical processing units.
\end{itemize}
From a quantitative perspective, the best-fit lens parameter values we obtained in our mock data tests were within $\sim 1\%$ of the true values, albeit with some bias at very high signal-to-noise ratios.

Since our fitting and posterior sampling is gradient-based, the computational cost of adding parameters to increase the realism of our model is low. For example, a prerequisite for applying our pipeline to real data is to model the light from the lens galaxy. We anticipate this could be done by adding an additional unlensed light source again modeled by our variational autoencoder. In future work, we will also explore more realistic lens models with additional components such as dark matter subhalos, a baryonic disk, line-of-sight halos, or a more complicated main lens model. For these very high-dimensional models, automated gradient-based inference techniques with favorable scaling behavior such as variational inference~\citep{2012arXiv1206.7051H,2016arXiv160100670B} could enable analysis of posterior distributions for hundreds or even thousands of parameters. Lastly, we expect that our pipeline's efficiency in analyzing large datasets could be improved by making technical changes so it can operate on batches of images in parallel, as is done when training convolutional neural networks.

While our pipeline is an early example of differentiable probabilistic programming, we anticipate this approach will enable other challenging and exciting data analyses in the future by leveraging the advantages of deep learning and physics modeling.

\section*{Acknowledgements}
We would like to thank Simona Vegetti and Rajat Thomas for helpful conversations. This work was carried out on the Dutch national e-infrastructure with the support of SURF Cooperative. We acknowledge funding from the Netherlands Organization for Scientific Research (NWO) through the VIDI research program ``Probing the Genesis of Dark Matter" (680-47-532).



\bibliographystyle{mnras}
\bibliography{references}



\appendix

\section{The main halo model}
\label{app:main_halo}

In this appendix, we describe how the contribution to the displacement field due to the main halo, the Singular Power-Law Ellipsoid (SPLE) profile, is implemented in the Lens Model block of the lensing pipeline shown in Fig.~\ref{fig:pipeline}. It is worth observing that, in the absence of an analytical expression for $\bm{\alpha}_{\rm mh}$, one has to numerically compute the integral in eq.~\eqref{eq:deflection}. However, this is not feasible in our framework because the numerical integration is not coded up in a autodifferentiable way. For this reason, the displacement field is instead determined by means of an interpolation of a precomputed numerical table of the corresponding integral, eq.~\eqref{eq:deflection}.

In case of a surface mass density profile with elliptical contours (like for example the SPLE profile), the two-dimensional integral of eq.~\eqref{eq:deflection} can be reduce to a simpler one-dimensional integral~\citep{Schramm, Barkana:1998qu}
\begin{eqnarray}
    \alpha_1\left(\theta_1,\theta_2\right) & =&  2 \theta_1 q \int_0^{\rho\left(\theta_1,\theta_2\right)} \frac{\rho'\,\kappa(\rho') \, \omega}{\theta_1^2 + \omega^4\,\theta_2^2} d\rho' \,, \label{eq:a_1} \\ 
    \alpha_2\left(\theta_1,\theta_2\right) & =&  2 \theta_2 q \int_0^{\rho\left(\theta_1,\theta_2\right)} \frac{\rho'\,\kappa(\rho') \, \omega^3}{\theta_1^2 + \omega^4\,\theta_2^2} d\rho' \,,
    \label{eq:a_2}
\end{eqnarray}
with
\begin{eqnarray}
    \omega^2 & = & \frac{\Delta + r^2 + \rho'^2 (1-q^2)}{\Delta + r^2 - \rho'^2 (1-q^2)} \,, \\
    \Delta^2 & = & \left[\rho'^2 (1-q^2) + \theta_2^2 - \theta_1^2\right]^2 + 4 \theta_1^2 \theta_2^2\,.
\end{eqnarray}
where $r^2 = \theta_1^2 + \theta_2^2$, $\rho\left(\theta_1,\theta_2\right)$ is defined in eq.~\eqref{eq:rho}, and the quantity $\kappa$ is the surface mass density $\Sigma$ in units of the critical density $\Sigma_{\rm cr}$. For a given profile, the integrals in eqs.~\eqref{eq:a_1} and~\eqref{eq:a_2} can be tabulated for different values of a subset of the lens parameters $\mathbf{\Theta}_{\rm lens}$. In particular, in case of the SPLE lens, the displacement field has been evaluated on a unit circle $(r=1)$, for which the coordinates are $\theta_1 = \cos \eta$ and $\theta_1 = \sin \eta$ with the angle $\eta$ being in the first quadrant ($0 \leq \eta \leq \pi/2$), for different values of the axis ratio $q$ in the interval $\left[0, 1\right]$ and the slope $\gamma$. The Einstein radius is instead fixed to be $r_{\rm Ein} = 1$. Such a procedure provided a three-dimensional table in the variables $\left\{\eta,\,q,\,\gamma\right\}$. This numerical table is then interpolated to compute the SPLE displacement field for any values of $q$ and $\gamma$, and at any position $(\theta_1, \theta_2)$. Each component of the displacement field is indeed given by
\begin{eqnarray}
\alpha_i \left(\theta_1, \theta_2\right) & = &  \alpha_i \left(|\theta_1|, |\theta_2|\right) \, {\rm sign}\left(\theta_i\right) \,, \\
\left.\alpha_i\left(\theta_1, \theta_2\right)\right|_{r} & = & r^{2-\gamma} \, \left.\alpha_i \left(\theta_1, \theta_2\right)\right|_{r=1}\,.
\end{eqnarray}
Moreover, in case of the SPLE profile, these two components show a simple scaling relation as a function of the Einstein radius. We have
\begin{equation}
\left.\alpha_i\left(\theta_1, \theta_2\right)\right|_{r_{\rm Ein}} = \left(\frac{1}{r_{\rm Ein}}\right)^{1-\gamma} \, \left.\alpha_i \left(\theta_1, \theta_2\right)\right|_{r_{\rm Ein}=1}
\end{equation}
It is worth noticing that the accuracy in computing the displacement field by means of this procedure depends on the size of the interpolation table, which can be defined without any constraint.

\section{ELBO derivation}
\label{app:elbo}

This appendix demonstrates one possible derivation of eq.~\eqref{eq:elbo_objective}. Consider a latent variable model defined by the joint probability distribution $p(x, z)$, where $x$ and $z$ are the observed and latent variables, respectively. We start by rewritting an expression for the log of the evidence:
\begin{eqnarray}
    \log p(x) &=& \log \int_z p(x, z) dz \\
     &=& \log \int_z p(z|x) \, \frac{p(x, z)}{p(z|x)} dz \,,
\end{eqnarray}
This can be recognized as an expectation value over $p(z|x)$:
\begin{align}
    \log p(x) &= \log \E_{p(z|x)} \left[ \frac{p(x, z)}{p(z|x)} \right].
\end{align}
By Jensen's inequality, which relates the expectation value of a convex function to that function applied to an expectation value, we have
\begin{align}
    \log \E_{p(z|x)} \left[ \frac{p(x,z)}{p(z|x)} \right] \geq \E_{p(z|x)} \left[ \log \left( \frac{p(x,z)}{p(z|x)} \right) \right],
\end{align}
and thus:
\begin{equation}
    \log p(x) \geq \mathbb{E}_{p(z|x)} \left[ \log \left( \frac{p(x,z)}{p(z|x)} \right) \right].
\end{equation}
Using the definition of the Kullback-Leibler divergence for continuous random variables
\begin{equation}
    D_{KL}[p || q] = -\mathbb{E}_{p(x)}\left[ \log \frac{q(x)}{p(x)} \right],
\end{equation}
this can be manipulated to yield
\begin{align}
    \log p(x) &\geq \E_{p(z|x)} \left[ \log \frac{p(x|z) p(z)}{p(z|x)} \right]\\
    &= \E_{p(z|x)} \left[ \log p(x|z) + \log \frac{p(z)}{p(z|x)} \right]\\
    &= \E_{p(z|x)} \left[ \log p(x|z) \right] - D_{KL}\left[ p(z|x) || p(z) \right]\\
    &\equiv \elbo(\theta, \phi; x).
\end{align}
The VAE training objective is obtained by substituting the approximate distributions $p(x | z) \to d_\theta(x | z)$ and $p(z | x) \to e_\phi(z | x)$ for the true ones.

\section{Optimizing the ELBO}
\label{app:reparameterization_trick}

Training the variational autoencoder requires taking the gradient of the ELBO for (batches of) training images $\{ x^{(i)} \}_{i=1}^N$ with respect to the encoder and decoder's parameters $\theta$ and $\phi$:
\begin{equation}
\begin{split}
    \nabla_{\theta, \phi} &\elbo\left( \theta, \phi; x^{(i)} \right)\\
    &= \nabla_{\theta, \phi}  \E_{e_\phi\left( z | x^{(i)} \right)} \left[ \log d_\theta\left( x^{(i)} | z \right) \right]\\
    &\hspace{1cm} - \nabla_{\theta, \phi} D_{KL}\left[ e_\phi\left( z | x^{(i)} \right) ||\,  p(z) \right].
\end{split}
\end{equation}
For the normal encoding distribution and latent space prior adopted in this work, the KL divergence term can be integrated analytically, which makes it simple to compute the second term on above. The first term is more challenging: while a Monte Carlo estimate of the derivative with respect to $\theta$ can be performed by sampling $\left\{ z^{(j)} \sim e_\phi\left( z | x^{(i)} \right) \right\}_{j=1}^M$, it is not obvious how to compute the derivatives of the sampled latent variable values with respect to $\phi$.

The solution is the reparameterization trick introduced in the two original papers on variational autoencoders~\citep{2013arXiv1312.6114K,2014arXiv1401.4082J}. The insight is that (assuming a normal encoding distribution) the randomness and $\phi$-dependent parts of the sampling process can be factored, allowing the sampled values to be written as
\begin{equation}
    z^{(j)} = \mu_e\left( x^{(i)}; \phi \right) + \epsilon^{(j)} \, \sigma_e\left( x^{(i)}; \phi \right),
\end{equation}
with $\epsilon^{(j)} \sim \mathcal{N}(0, I)$. These can be used to construct the following Monte Carlo gradient estimator by treating the $\epsilon^{(j)}$ values as constants for the optimization epoch:
\begin{equation}
\begin{split}
    \nabla_{\theta, \phi} &\E_{e_\phi\left( z | x^{(i)} \right)} \left[ \log d_\theta\left( x^{(i)} | z \right) \right]\\
    &\approx \frac{1}{M} \sum_{j=1}^M \log d_\theta\left( x^{(i)} | z^{(j)} \right).
\end{split}
\end{equation}
This estimator is stable; we set $M=10$ in our work by using batches of $32$ training images.

\section{Variational Autoencoder Architecture}
\label{app:vae_architecture}

The architectural details of the encoder and decoder networks of our variational autoencoder are presented in Tables~\ref{tab:encoder} and \ref{tab:decoder}, respectively. All weights were initialized to $0.02$. The biases in the final linear layers of the encoder were initialized to $0$.

We tried several experiments to see whether we could improve upon this VAE design. For example, since the decoder can have trouble saturating the final $\tanh$ activation, we tried exchanging this for a LeakyReLU, as well as removing it altogether. We also tested 32 and 128 latent-space dimensions. The former lead to blurry reconstructions while the later yielded little improvement relative to 64 latent-space dimension. \citet{2019arXiv190305789D} analytically demonstrated that making the hyperparameter $\sigma_d$ a trainable parameter should lead to sharper reconstructions. We did not find this to be the case, and instead found that $\sigma_d$ converged to roughly the value we selected by hand based on the signal-to-noise ratio of the training data. Finally, we experimented with a residual network-based architecture, as was used in \citet{2019arXiv190305789D}, which did not yield any improvement relative to the architecture we eventually selected.

\begin{table}
    \centering
    \begin{tabular}{c c}
        \hline\hline
        \multicolumn{2}{c}{Conv2d(1, 64, 4, 2, 1)}\\
        \multicolumn{2}{c}{LeakyReLU(0.2)}\\
        \hline
        \multicolumn{2}{c}{Conv2d(64, 128, 4, 2, 1)}\\
        \multicolumn{2}{c}{BatchNorm2d(128)}\\
        \multicolumn{2}{c}{LeakyReLU(0.2)}\\
        \hline
        \multicolumn{2}{c}{Conv2d(128, 256, 4, 2, 1)}\\
        \multicolumn{2}{c}{BatchNorm2d(256)}\\
        \multicolumn{2}{c}{LeakyReLU(0.2)}\\
        \hline
        \multicolumn{2}{c}{Conv2d(256, 512, 4, 2, 1)}\\
        \multicolumn{2}{c}{BatchNorm2d(512)}\\
        \multicolumn{2}{c}{LeakyReLU(0.2)}\\
        \hline
        \multicolumn{2}{c}{Conv2d(512, 4096, 4, 1, 0)}\\
        \multicolumn{2}{c}{LeakyReLU(0.2)}\\
        \hline
        \multirow{2}{*}{Linear(4096, 64)} & Linear(4096, 64)\\
                                          & Exp\\
        \hline
        $\mu_e(x)$ & $\sigma_e(x)$\\
        \hline\hline
    \end{tabular}
    \caption{\textbf{Encoder neural network architecture.} The notation uses the same conventions as pytorch. The arguments of Conv2d indicate the number of input channels, number of output channels, kernel size, stride and zero padding; all convolutions are unbiased. The LeakyReLU argument is slope for inputs less than 0. The BatchNorm2d argument is the number of input channels. The output of the last convolutional block is flattened before being passed to the two separate Linear layers to produce the mean and standard deviation of the encoding distribution. Linear layers' arguments show the number of input and output channels.}
    \label{tab:encoder}
\end{table}

\begin{table}
    \centering
    \begin{tabular}{c}
        \hline\hline
        ConvTranspose2d(64, 512, 4, 1, 0)\\
        BatchNorm2d(512)\\
        ReLU\\
        \hline
        ConvTranspose2d(512, 256, 4, 2, 1)\\
        BatchNorm2d(256)\\
        ReLU\\
        \hline
        ConvTranspose2d(256, 128, 4, 2, 1)\\
        BatchNorm2d(128)\\
        ReLU\\
        \hline
        ConvTranspose2d(128, 64, 4, 2, 1)\\
        BatchNorm2d(64)\\
        ReLU\\
        \hline
        ConvTranspose2d(64, 1, 4, 2, 1)\\
        Tanh\\
        \hline
        $\mu_d(z)$\\
        \hline\hline
    \end{tabular}
    \caption{\textbf{Decoder neural network architecture.} The notation is described in the caption of Table~\ref{tab:encoder}, and is the same for Conv2d and ConvTranspose2d. The input vector $\mathbf{z}$ is reshaped to have 64 channels and spatial dimensions equal to 1 along both axes.}
    \label{tab:decoder}
\end{table}

\bsp	
\label{lastpage}
\end{document}